# Intersection of triangles in space based on cutting off segments

Irina Bolodurina[1][0000-0003-0096-2587], Georgii Nigmatulin[1][0000-0001-8935-1146] and Denis Parfenov[1][0000-0002-1146-1270]

[1] Orenburg State University, Orenburg, Russia
`prmat@mail.osu.ru`

**Abstract.** The article proposes a new method for finding the triangle-triangle intersection in 3D space, based on the use of computer graphics algorithms – cutting off segments on the plane when moving and rotating the beginning of the coordinate axes of space. This method is obtained by synthesis of two methods of cutting off segments on the plane – Cohen — Sutherland algorithm and FC-algorithm. In the proposed method, the problem of triangle-triangle intersection in 3D space is reduced to a simpler and less resource-intensive cut – off problem on the plane. The main feature of the method is the developed scheme of coding the points of the cut-off in relation to the triangle segment plane. This scheme allows you to get rid of a large number of costly calculations. In the article the cases of intersection of triangles at parallelism, intersection and coincidence of planes of triangles are considered. The proposed method can be used in solving the problem of tetrahedron intersection, using the finite element method, as well as in image processing.

**Keywords:** triangle-triangle intersection, computer graphics, cut-off algorithm, Cohen — Sutherland algorithm, FC-algorithm.

## 1 Introduction

The triangle-to-triangle intersection test is a basic component of all collision detection data structures and algorithms. Most collision detection algorithms try to minimize the number of primitive-primitive intersections that have to be computed. Still, a fast and reliable method for computing the primitive-primitive intersection is desired. Since rendering hardware is often targeted for triangles, the primitives in collision detection algorithms are often triangles as well.

Several solutions exist to test the intersection between three-dimensional triangles. In paper [1] authors consider technique solve the basic sets of linear equations associated with the problem and exploits the strong relations between these sets to speed up their solution. Held's method [2] determining on which side of the supporting plane the vertices lie and then find the intersections using two-dimensional triangle tests using projection onto a convenient plane. Moller [3] resents a method, along with some optimizations, for computing whether or not two triangles intersect. Elsheikh's [4] algorithm is based on an extensive set of triangle - edge intersection cases, com-bined with an intersection curve tracking method. In this paper, we try to describe a new method for determining if two triangles intersect which allows getting rid of a large number of costly calculations.

## 2    Setting a task

Consider the problem of finding the intersection of two triangles ΔABC and ΔDEF in 3-dimensional space, each of which is given by 3 points such ΔABC: $A(x_1, y_1, z_1)$, $B(x_2, y_2, z_2)$, $C(x_3, y_3, z_3)$ and ΔDEF: $D(x_4, y_4, z_4)$, $E(x_5, y_5, z_5)$, $F(x_6, y_6, z_6)$. Find the intersection of two triangles ΔABC andΔDEF.

The reciprocal arrangement of triangles can be as follows (see **Error! Reference source not found.**).

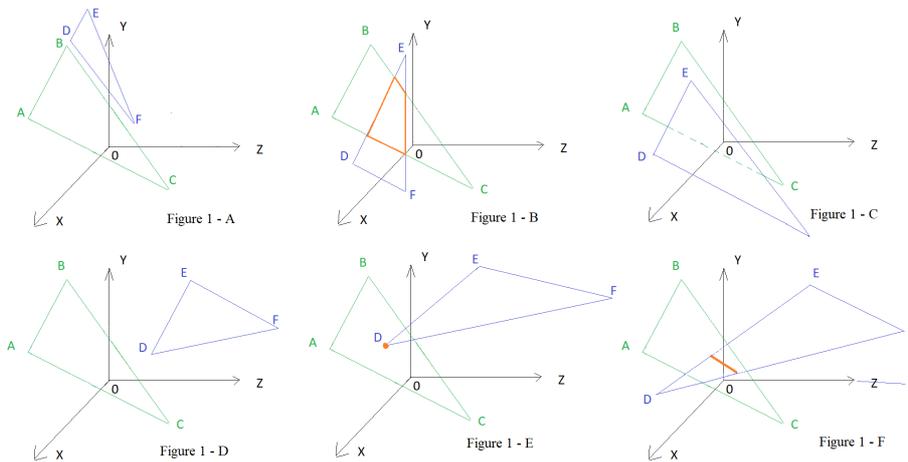

**Fig. 1.** The reciprocal arrangement of the triangles of ΔABC и ΔDEF.

1. Triangles of ΔABC and ΔDEF do not have intersection points in space - the plane triangles are the same, but there are no common points. (Fig. 1 - B).
2. Triangles of ΔABC and ΔDEF have a contour of intersection in space - the planes of the triangles are the same and there are common points. (Fig. 1 - D).
3. Triangles of ΔABC and ΔDEF have no intersection points in space - the planes of the triangles are parallel. (Fig. 1 - C)
4. Triangles of ΔABC and ΔDEF do not have intersection points in space - the plane triangles intersect, but the triangles do not. (Fig. 1 - A)
5. Triangles of ΔABC and ΔDEF have 1 point of intersection in space - the planes of triangles intersect. (Fig. 1 - E)
6. Triangles of ΔABC and ΔDEF have 2 points of intersection in space - the plane triangles intersect. (Fig. 1 - F)

## 3 Intersection triangles in intersecting planes

### 3.1 Displaying the dots of the ΔDEF triangle on the plane of the triangle ΔABC

When the planes of the triangles intersect, you need to find the dots belonging to one of the planes of the triangle and at the same time to the sides of the other triangle. To do this, you need to fix one of the planes, let it be the plane of the triangle ΔABC - and find the intersection points of each side of the triangle ΔDEF with this plane.

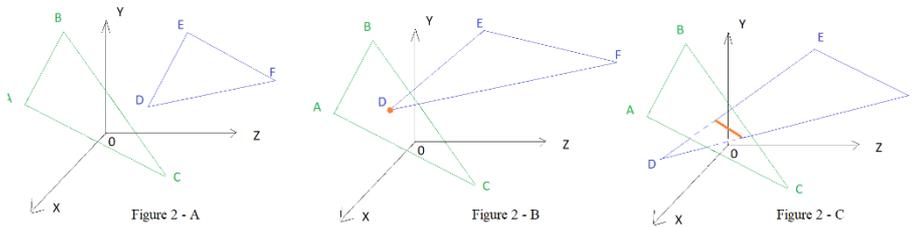

**Fig. 2.** The reciprocal arrangement of the triangles of ABC and def in non-parallel planes.

Deciding the system (1) consisting of a straight-line equation set on the two points of the side of the triangle ΔDEF, and the plane equation $P_1$, we will find the intersection of the side of the triangle with the plane $P_1$.

$$\begin{cases} \frac{x-x_1}{x_2-x_1} = \frac{y-y_1}{y_2-y_1} = \frac{z-z_1}{z_2-z_1} \quad - \quad straight-line\ equation \\ Q_1 x + W_1 y + U_1 z + R_1 = 0 \quad - \quad plane\ equation \end{cases} \quad (1)$$

The best way to solve this system by using parametric equations of the straight-line;

$$\begin{cases} x = x_1 + t(x_2 - x_1) \\ y = y_1 + t(y_2 - y_1) \\ z = z_1 + t(z_2 - z_1) \end{cases}, next \quad \begin{cases} x = x_1 + t*m \\ y = y_1 + t*n \\ z = z_1 + t*p \end{cases} \quad (2)$$

Each value of t corresponds to a point of the line. You must select a value t at which the point of the line will lie on the plane $P_1$. Substituting the data x, y, z in the equation of the plane $P_1$ and make the transformations, you can deduce the value of the variable t.

$$t = -\frac{Ax_1 + By_1 + Cz_1 + D}{A*m + B*n + C*p} \quad (3)$$

Substituting the found t value into the parametric equations of the straight (2), we will find the coordinates of the point of intersection straight to the plane.

Next, you need to check whether the found point belongs to the segment that makes up the side of the triangle. To do this, it is enough to substitute the found values in the original side equation of the side of the side.

$$\frac{x-x_1}{x_2-x_1} = \frac{y-y_1}{y_2-y_1} = \frac{z-z_1}{z_2-z_1} \quad - \quad side\ of\ the\ triangle\ \Delta DEF \tag{4}$$

If these equalities are executed, the point belongs to the side and intersects the plane. This procedure should be applied to each side of the triangle. The result of the action will be either 2 points, or 1 point, or 0. If no points of intersection between the plane and the sides of the triangles are found (Figure 2 - A), we can conclude that the triangles ΔABC and ΔDEF do not intersect. If 1 point is found (Figure 2 - B), then the triangles ΔDEF and ΔABC touch each other (have one point of intersection), and the point under consideration is the point of contact. If 2 points are found (Figure 2 - C), it is necessary to continue the research to find the direct intersection of two triangles.

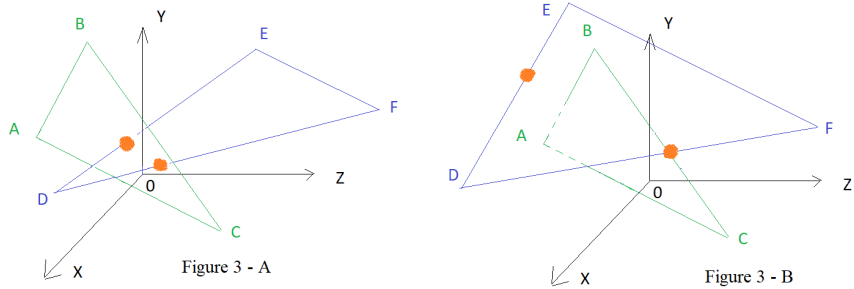

Figure 3 - A    Figure 3 - B

**Fig. 3.** The reciprocal arrangement of the triangles is ΔABC when 2 points appeared on the plane of the triangle.

Next, let's consider the situation when, after displaying the dots of the triangle, 2 points appeared on the plane of the triangle (Figure 3 - A) and (Figure 3 - B).

### 3.2 Moving and rotation the beginning of the coordinate axes of space

After displaying the dots of the triangle ΔDEF on the plane of the triangle ΔABC, 2 points appeared on the P2 plane. To accomplish this task, we will use the cut-off algorithm described below, which will ensure that the program is highly efficient. But the main condition of this algorithm is that the cut-off should take place in a two-dimensional plane, so it is necessary to move and turn the beginning of the coordinate systems so that the value of one coordinate of all points is zero. This is possible because all 5 points lie on the same plane. Let's move the beginning of the coordinate systems so that the coordinates of point A become equal to 1 excluding x coordinate equating this to 0 - A(1,1,0).

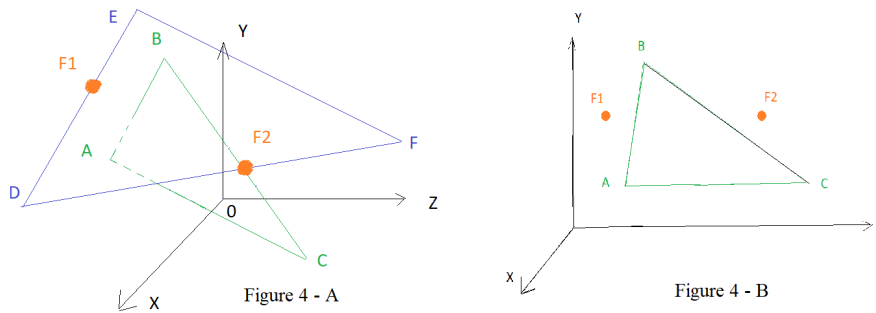

**Fig. 4.** Move and turn the beginning of the coordinate axles of space.

As you can see, Figure 4 - B shows the triangle of the ABC and the 2 points of the triangle ΔDEF - F1 and F2. We can say that the points A, B, C, F1, F2 are located on the 2-dimensional plane Z0Y.

### 3.3 Dot coding

To find the intersection of triangles, an algorithm - "Cutting the triangle" - obtained by the synthesis of two methods of cutting off segments - the algorithm Cohen - Sutherland and FC- algorithm, in which instead of a rectangular cutting area the triangle is used.

This algorithm allows you to quickly identify segments that can be either accepted or discarded entirely. The calculation of intersections is required when the segment does not fall into any of these classes. This algorithm is particularly effective in two extreme cases, such as

─ Most instances are contained entirely in a large window;
─ Most instances lie entirely outside the relatively small window.

The triangle cut-off algorithm divides the plane into 7 straight parts that form the sides of the triangle. Each of the 7 parts is assigned a three-bit code in a way that is shown on Figure 5.

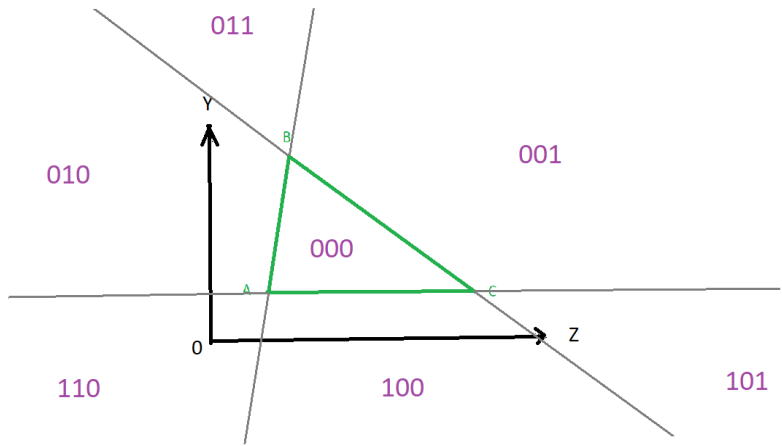

**Fig. 5.** Coding parts of a plane separated by the sides of a triangle.

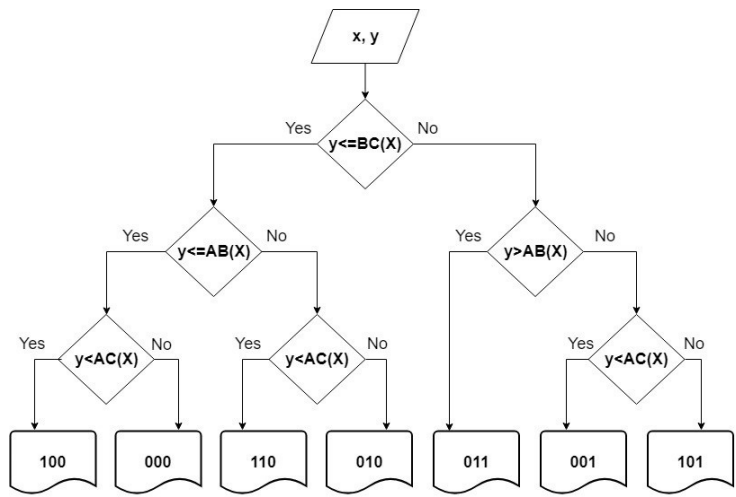

**Fig. 6.** Point coding algorithm.

To determine whether a segment lies entirely inside the window or entirely outside the window, do the following:

1. If the codes of both ends of the segment are 0, the segment is entirely inside the window, clipping is not necessary, the segment is accepted as <u>trivially visible</u>. For example, the segment N1N2 (Figure 7);
2. If the logical conjunction of codes of both ends of the segment are not zero, then the segment is completely out of the window, clipping is not necessary, the segment is discarded as <u>trivially invisible.</u> For example, the segment N3N4 (Figure 7);
3. If the logical conjunction and codes of both ends of the segment are zero, then the segment <u>is suspicious</u>, it can be partially visible (segments N5N6, N7N8, N9N10) or completely invisible (section 11N12); it needs to determine the coordinates of the intersections with the sides of the triangle and for each received part to determine trivial visibility or invisibility.

If only 1 point of touch and plane is displayed on the plane of the triangle, the conclusion about its trivial visibility is also made on the basis of codes - if the point code is "000", the point is accepted, in otherwise, is rejected.

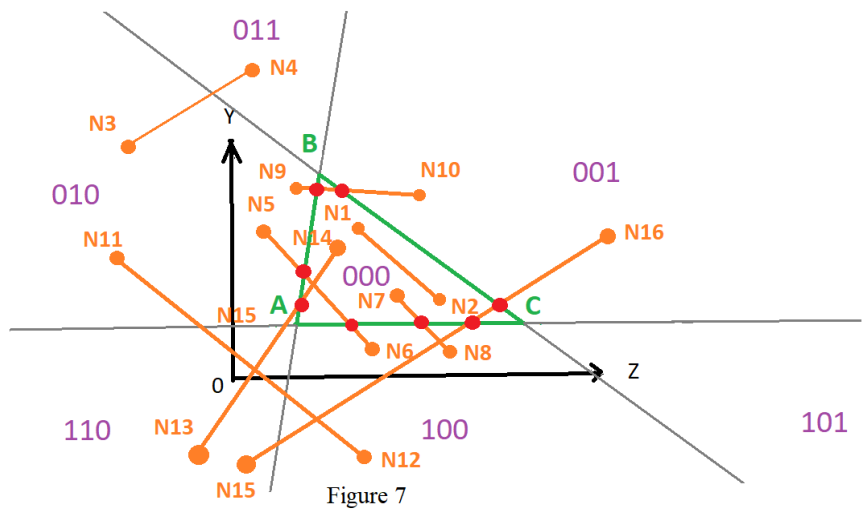

**Fig. 7.** Possible points and segment positions.

It is these rules that allow you to quickly decide whether to work with the segment in question or not. After matching the points of one of the 7 codes, it is necessary to start cutting off the segment by considering one of the points of the displayed triangle ΔDEF. There are three different situations.

- The point in question is on the outside, and the area to which it <u>relates concerns the side of the triangle</u> ΔABC. For Example, N5, N9, N10 (Figure 7)
- The point in question is located outside and the area to which it belongs <u>relates to the vertices of the triangle</u> ΔABC. For Example, N4, N13 (Figure 7)
- The point in question is inside <u>the triangle.</u> For Example, N1, N2, N7 (Figure 7).

### 3.4   The «entry» and «exit» of the intersection of the section and the triangle

Coding points on the above diagram allows you not only to determine whether it is possible to intersect a segment with a triangle, but also signals which side of the segment can entry and exit to intersect the triangle. If the point belongs to the area touching the sides of the triangle ΔABC, then only one option of the beginning of the intersection is possible: with the side with which this area of coding touches. For example, segments leaving points N5 and N9 with the code "010" can intersect only the side AB (segments N5N6, N9N10). If they do not intersect this side, then there is no intersection with the triangle. Suppose that the intersection of the segment and the side occurred, then, if the second point of the segment is outside the triangle, then this segment can pass through the other two sides and requires finding the intersection of this segment and the sides of the triangle. For example, the intersection of segment N5N6 and side AC, segment N9N10 and side BC.

In the case of points belonging to areas relating to the vertices of the triangle ΔABC, then there are 2 options for the beginning of the intersection with each of the two sides of the triangle forming this area of coding. For example, segments coming out of points N13 and N15 with code "110" can intersect both the AB side (segment N13N14) and the AC side (segment N15N16). If the segment does not intersect these sides, then there is no intersection with the triangle. Suppose that the intersection of a segment and one of the sides occurred, then if the second point of the segment is outside the triangle, then this segment will definitely pass through the third side, which is not involved in the formation of the coding region of the first point, for example, the side BC for the segment N15N16.

If the point is inside the ΔABC triangle and has the code "000", you must check the intersection of the segment coming out of that point with each side of the triangle until the intersection is found.

All the rules by which the entry and exit (if any) of the intersections are carried out are presented in tables 1, 2, 3.

**Table 1.** Entry and exit for entry points from the area touching the sides of the triangle..

| Entry and exit for entry points from the area with the sides of the triangle | | | |
|---|---|---|---|
| Entry point code | Entry party | Exit point code | Exit party (if there is) |
| 010 | AB | 000 | - |
| 010 | AB | 100 | AC |
| 010 | AB | 001 | BC |
| 010 | AB | 101 | AC or BC |
| 100 | AC | 000 | - |
| 100 | AC | 010 | AB |
| 100 | AC | 001 | BC |
| 100 | AC | 011 | AB or BC |
| 001 | BC | 000 | - |
| 001 | BC | 100 | AC |
| 001 | BC | 010 | AB |
| 001 | BC | 110 | AB or AC |

**Table 2.** Entrance and exit for entry points from the area dealing with the tops of the triangle.

| Entry and exit for entry points from the area dealing with the vertices of the triangle |
|---|

| Entry point code | Entry party | Exit point code | Exit party (if there is) |
| --- | --- | --- | --- |
| 110 | AB or AC | 000 | - |
| 110 | AB or AC | 001 | BC |
| 011 | AB or BC | 000 | - |
| 011 | AB or BC | 100 | AC |
| 101 | BC or AC | 000 | - |
| 101 | BC or AC | 010 | AB |

**Table 3.** The entrance and exit for the entrance points lying inside the triangle.

| Entrance and exit for entry points lying inside the triangle | | | |
| --- | --- | --- | --- |
| Entry point code | Entry party | Exit point code | Exit party (if there is) |
| 000 | - | 000 | - |
| 000 | - | 110 | AB or AC |
| 000 | - | 010 | AB |
| 000 | - | 011 | AB or AC |
| 000 | - | 001 | BC |
| 000 | - | 101 | BC or AC |
| 000 | - | 100 | AC |

### 3.5 Key steps in the triangle cut-off algorithm

Once the section intersection operation has been defined, you can formulate the main stages of the triangle cut-off algorithm.

1. Detection the areas where the dots are located.
2. A selection of sides with which to check the intersection of the segment and the triangle. (Choosing parties according to the rules from tables 1, 2, 3).
3. Finding the intersection of the segment and the sides of the triangle on the plane.
4. Return the coordinate axes to the starting position (moving and turning).

Note that the result of the algorithm will be the coordinates of one or two points of the triangle $\Delta ABC$. If the answer - the coordinates of the two points, then in order to find a common equation straight - it is enough to take the vector work of points in the homogeneous coordinates- Intersect $(y_1, z_1, 0)$, $(y_2, z_2, 0)$.

## 4 Intersection triangles in concurring planes

There are two options for the location of triangles with matching planes, such as the same planes.

─ triangles $\Delta ABC$ and $\Delta DEF$ have no intersection points(Figure 8 - A);
─ triangles $\Delta ABC$ and $\Delta DEF$ have a circuit intersection (Figure 8 - B).

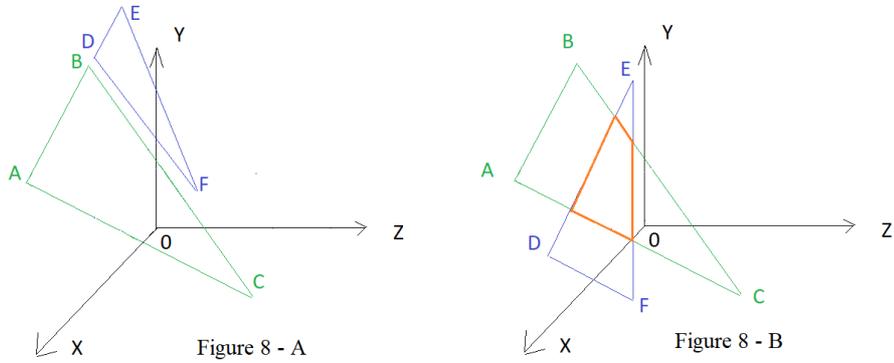

**Fig. 8.** The location of triangles with matching planes.

Since the triangles ΔABC and ΔDEF are on the same plane, it becomes possible to move and rotate the beginning of the coordinate axes of space in such a way that one coordinate of all points becomes zero. This will allow analysis of the intersections of the triangles methods of clipping a polygon in the plane. Let us use the same displacement and rotation of the plane as in point 2.2 - Displacement and rotation of the beginning of the coordinate axes of space.

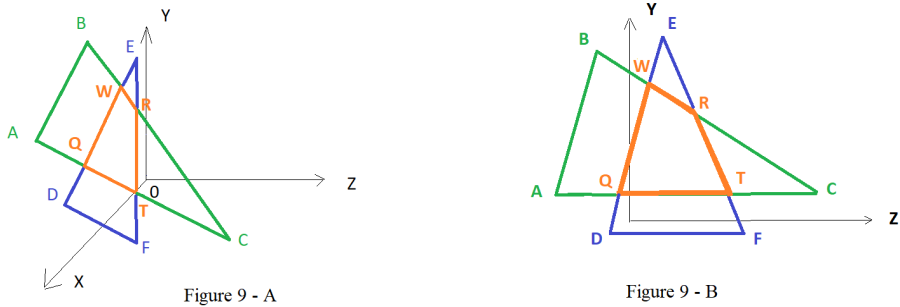

**Fig. 9.** Move and turn the beginning of the coordinate.

We will call the triangle ΔABC used as the clipping window a clipper or a window, and the triangle that is clipped is clipped ΔDEF. In principle, this problem can be solved using the above algorithm of cutting lines by a triangle, if we consider a triangle simply as a set of vectors, and not as a solid region. In this case, the vectors that make up the triangle are simply sequentially cut off by the sides of the window. If a closed polygon is to be obtained as a result of clipping, a vector connecting the last visible point with the point of intersection with the window is formed. In this case, we need to find a closed polygon, which is the intersection of triangles. Use the algorithm of polygon clipping Weiler-Atherton where consider a special case of polygon - triangle.

## 4.1 Weiler-Aerton Triangle Cut-off Algorithm

Let's say that each of the triangles is set by a list of tops of ΔABC (A, B, C), ΔDEF (D, E, F). Two types of dots occur between the boundaries of the compartmental polygon and the windows.

— input points when the oriented edge of the clipped polygon enters the window.
— output points when the edge of the clipped polygon goes from the inside to the outside of the window.

The general algorithm of the Weiler-Atherton algorithm to determine the portion of the intercept of the polygon, hitting the window, the next:

1. Lists of vertices of the cut-off triangle and the window are constructed.
2. All points of intersection are found. To perform this action encoding is used and the cutting off of segments of a triangle, are discussed in paragraphs 3.3 and 3.4. In this calculation, touches are not considered to be an intersection, that is, when the vertex or edge of the clipped triangle is incident or coincides with the side of the window.
3. Lists of coordinates of the vertices of the cut-off triangle and windows are supplemented by new tops - coordinates of the points of intersection. And if the point of intersection $F_k$ is on the edge connecting the tops $R_i$, $R_j$ of triangle, the sequence of points turns into a sequence $R_i$, $F_k$, $R_j$. At the same time, two-way connections between the same intersection points in the lists of the tops of the cut-off triangle and window are established. The input and output intersection points form separate sublists of the input and output points in the vertex lists.
4. Definition of the part of the processed triangle which has got to the window. This is done by this algorithm:

— If the list of input intersection points is not exhausted, select the next input point;
— Move along the vertices of the cut-off triangle until the next intersection point is found; all passed points, not including the interrupted view, are entered in the result.
— Using the two-way connection of the intersection points, switch to view the list of vertices of the window.
— Move along the vertices of the window until the next intersection point is detected; all passed points, not including the last one that interrupted the view, are recorded in the result.
— Using the two-way connection of the intersection points, switch to the list of vertices of the processed triangle.

These steps repeat until the start point of the clipping is part of the intercept of the polygon, hitting the window closed.

Let's demonstrate the work of the Weiler-Azerton algorithm by the example of the intersection of two triangles ΔABC and ΔDEF in the plane, where the triangle ΔABC is the cut-off window, and the triangle ΔDEF is the cut-off triangle (Figure 10 - A).

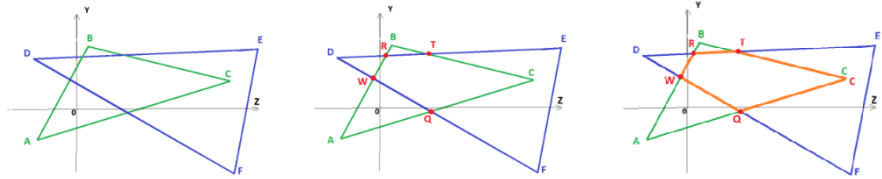

**Fig. 10.** Weiler-Atheerton algorithm.

1. Let's build a list of vertices of the triangles ΔABC, used as cut-off window clipping, and the intercept of the triangle ΔDEF (figure 10 - a).

Table 4. A list of the tops of the triangles of DEF and ABC.

| Cut off - ΔDEF | The Cut-Off - ABC |
|---|---|
| D | A |
| E | B |
| F | C |
| D | A |

2. Let's find all points of intersection of segments and triangle ΔABC – points Q, W, R, T. For effective finding of intersections we use coding and clipping described in points 3.3 and 3.4. (Figure 10 - B)
3. We will supplement the lists of coordinates of the tops of the cut-off triangle and the window coordinates of the intersection points. And we will establish two-way connections between the intersection points of the same name in the lists of the tops of the cut-off triangle and the window (dots with two-way ties are underlined and bold).

Table 5. Supplemented by the points of intersection of a list of vertices of the triangles ΔABC and ΔDEF.

| Cut off - ΔDEF | The Cut-Off - ABC |
|---|---|
| D | A |
| **R** | **W** |
| **T** | **R** |
| E | B |
| F | **T** |
| **Q** | C |
| **W** | **Q** |
| D | A |

4. On the basis of the obtained table, we make a contour along which the intersection of the triangles ΔABC and ΔDEF will pass. Initially, the first point of the triangle ΔDEF, which intersects the triangle ΔABC – point "R", is selected. Next, move along the vertices of the triangle cut off until you find the next intersection point – point "T". Using the two-way connection of the intersection points, switch to view the list of vertices of the window. We move along the tops of the window until we find the next intersection point – "Q". Note that the point "C" - the top of the cutoff ΔABC has been passed, it belongs to the contour as well as to the previous points. Using the two-way connection of the intersection points, switch to the list of vertices of the processed triangle. We move along the vertices of the cut-off triangle until we find the next intersection point – "W". Using the two-way connection of the intersection points, switch to view the list of vertices of the window. We move along the tops of the window until we find the next intersection point – "R". Since the "R" point is the starting point from which began the construction of a cut-off stop the search cut-off.

Let's see how you can build a contour on the table:

| Cut off - ΔDEF | Cut-off value - ΔABC |
|---|---|
| D | A |
| **R** | **W** |
| **T** | **R** |
| E | B |
| F | **T** |
| **Q** | C |
| **W** | **Q** |
| D | A |

**Fig. 11.** Construction of a contour of intersection of triangles on the list of vertices and points of intersection.

Built route «R» - «T» - «C» - «Q» - «W» - «R». (Figure 10 - C).

## 5      Conclusion

In this article we have considered various cases of intersection of triangles according the conditions of parallelism, intersection and coincidence of planes of triangles. A new method for finding the intersection of triangles on a plane, called "triangle Clipping", was proposed. This method was obtained by synthesis of two methods of cutoff – Cohen — Sutherland algorithm and FC-algorithm. The main feature of the proposed method is that a triangle is used instead of a rectangular clipping area. The proposed intersection rules and point encoding eliminate a large number of costly calculations. The main value of this method is to reduce the amount of computation. This can find application in using the finite element method and solution of problem of image processing. The use of fast modern algorithms for image processing, allows you to improve the quality of many intelligent transport systems. Thus, the speed of decision - making increases-which is important in the design of self-driving vehicles and traffic flow control, based on the analysis of traffic intensity.